\documentclass[a4paper]{article}

\usepackage{17lomcon}        
\usepackage{cite}             
\usepackage{bm,amsmath,amssymb}

\begin{document}


\title{ELECTROWEAK INTERACTION OF PARTICLES WITH ACCELERATED MATTER AND ASTROPHYSICAL APPLICATIONS}

\author{Maxim Dvornikov\email{maxdvo@izmiran.ru}}

\affiliation{Pushkov Institute of Terrestrial Magnetism, Ionosphere
and Radiowave Propagation (IZMIRAN),
142190 Troitsk, Moscow, Russia}

\affiliation{Physics Faculty, National Research Tomsk State University,
36 Lenin Avenue, 634050 Tomsk, Russia}


\date{}
\maketitle


\begin{abstract}
The description of physical processes in accelerated frames opens a window to numerous new phenomena. One can encounter these effects both in the subatomic world and on a macroscale. In the present work we review our recent results on the study of the electroweak interaction of particles with an accelerated background matter. In our analysis we choose the noninertial comoving frame, where matter is at rest. Our study is based on the solution of the Dirac equation, which exactly takes into account both the interaction with matter and the nonintertial effects. First, we study the interaction of ultrarelativistic neutrinos, electrons and quarks with the rotating matter. We consider the influence of the matter rotation on the resonance in neutrino oscillations and the generation of anomalous electric current of charged particles along the rotation axis. Then, we study the creation of neutrino-antineutrino pairs in a linearly accelerated matter. The applications of the obtained results for elementary particle physics and astrophysics are discussed.
\end{abstract}

Nowadays it is understood that noninertial effects are important in various areas of modern science such as elementary particles physics, general and special relativity, as well as condensed matter physics~\cite{review}. Recently in Refs.~\cite{Dvo14,Dvo15a,Dvo15b} it was realized that the electroweak interaction of particles with accelerated matter leads to interesting applications in physics and astrophysics. In those works, the treatment of the particle evolution was made in the comoving frame, where matter is at rest, with the noninertial effects being accounted for exactly. In the present work we review our recent results on the particle interaction with accelerated matter.

Our study of the fermion propagation in an accelerated matter is based on the Dirac equation in a comoving frame. In this situation one can unambiguously define the interaction with background matter. It is known that the motion in a noninertial frame is equivalent to the interaction with an effective gravitational field having the metric tensor $g_{\mu\nu}$. The Dirac equation for the particle bispinor $\psi$ in curved space-time has the form~\cite{Dvo15a},
\begin{equation}\label{eq:Depsicurv}
  \left[
    \mathrm{i}\gamma^{\mu}(x)\nabla_{\mu}-m
  \right]
  \psi =
  \gamma_{0}(x)
  \left\{
    \frac{V_{\mathrm{L}}}{2}
    \left[
      1-\gamma^{5}(x)
    \right] +
    \frac{V_{\mathrm{R}}}{2}
    \left[
      1+\gamma^{5}(x)
    \right]
  \right\}
  \psi,
\end{equation}
where $\gamma^{\mu}(x)$ are the coordinate dependent Dirac matrices, $\nabla_{\mu}=\partial_{\mu}+\Gamma_{\mu}$ is the covariant derivative,
$\Gamma_{\mu}$ is the spin connection, $m$ is the particle mass, $V_{\mathrm{L,R}} \sim G_\mathrm{F} n_\mathrm{eff}$ are the effective potentials of the interaction of left and right chiral projections with background matter,  $G_\mathrm{F}$ is the Fermi constant, $n_\mathrm{eff}$ is the effective density of background particles, $\gamma^{5}(x) = -(\mathrm{i}/4!) E^{\mu\nu\alpha\beta} \gamma_{\mu}(x) \gamma_{\nu}(x) \gamma_{\alpha}(x) \gamma_{\beta}(x)$,
$E^{\mu\nu\alpha\beta} = \varepsilon^{\mu\nu\alpha\beta} / \sqrt{-g}$
is the covariant antisymmetric tensor in curved space-time, and $g=\det(g_{\mu\nu})$. 

In Ref.~\cite{Dvo14} we found the solution of Eq.~(\ref{eq:Depsicurv}) for an ultrarelativistic neutrino moving in a rotating matter. Note that, in case of neutrinos, we should set $V_{\mathrm{R}}=0$. Choosing the appropriate vierbien vectors, we obtained $\psi$, which is expressed in terms of the Laguerre functions. Then we generalized our result to include different neutrino eigenstates and mixing between them. We obtained that the resonance condition is shifted by the matter rotation contrary to our previous claim in Ref.~\cite{Dvo10}. This effect can have the implication for the explanation of great linear velocities of pulsars since there is a correlation between the linear and angular velocities of a pulsar~\cite{Joh05}.

In Ref.~\cite{Dvo15a}, we obtained the solution of Eq.~\eqref{eq:Depsicurv} for ultrarelaticistic electroweakly interacting electrons and quarks in the rotating matter. Using this solution we derived the nonzero electric current along the rotation axis in the form,
\begin{equation}\label{eq:elcurr}
  \mathbf{J} =
  \frac{q\bm{\omega}}{\pi}
  \left(
    V_{\mathrm{R}}\mu_{\mathrm{R}}-V_{\mathrm{L}}\mu_{\mathrm{L}}
  \right),
\end{equation}
where $\bm{\omega}$ is the angular velocity, $q$ is the electric charge (including the sign) of a test fermion, and $\mu_{\mathrm{R,L}}$ are the chemical potentials of right and left fermions. The existence of the nonzero current in Eq.~\eqref{eq:elcurr} is attributed in Ref.~\cite{Dvo15a} to the new \emph{galvano-rotational effect} (GRE). GRE is analogous to the chiral vortical effect~\cite{Vil79}, in which the induced current is $\mathbf{J} \sim \bm{\omega} (\mu_{\mathrm{L}}^2 - \mu_{\mathrm{R}}^2)$. However, in the later case the current is vanishing in the equilibrium at $\mu_{\mathrm{L}} = \mu_{\mathrm{R}}$, whereas $\mathbf{J}$ in Eq.~\eqref{eq:elcurr} is nonzero in this situation.

GRE can be used for the generation of a toroidal magnetic field (TMF) in neutron and quark/hybrid stars. It is well known that, in a star, a purely poloidal magnetic field, which is observed by astronomers, is unstable. A toroidal component, which lays inside a star and can be of the same magnitude as a poloidal one, is required. In Ref.~\cite{Dvo15a} we estimated the strength of TMF generated owing to GRE as $B_\mathrm{tor}\sim |\mathbf{J}|R$, where $R\sim 10\thinspace\text{km}$ is the star radius.
Using the characteristics of the background matter in a compact star, one gets that $B_\mathrm{tor} \sim 10^8\thinspace\text{G}$ can be generated~\cite{Dvo15a}. This TMF strength is comparable with the observed magnetic fields in old millisecond pulsars~\cite{PhiKul94}.

In Ref.~\cite{Dvo15b} we solved Eq.~\eqref{eq:Depsicurv} for an ultrarelativistic neutrino interacting with a linearly accelerated matter. In this case $\psi$ is expressed via the Whittaker functions. The obtained solution turned out to reveal the instability of the neutrino vacuum leading to the creation of the neutrino-antineutrino ($\nu\bar{\nu}$) pairs. This phenomenon is analogous to the well known Unruh effect~\cite{CriHigMat08} consisting in the emission of the thermal radiation by an accelerated particle, with the effective temperature $T_\mathrm{eff} = a/2\pi$, where $a$ is the particle acceleration.

Requiring that the probability of the creation of $\nu\bar{\nu}$ pairs is not suppressed, in Ref.~\cite{Dvo15b}, we obtained the upper bound on the neutrino mass,
\begin{equation}\label{eq:masslim}
  m \lesssim m_{\mathrm{cr}},
  \quad
  m_{\mathrm{cr}}=2\sqrt{\frac{|V_\mathrm{L}|a}{\pi}}.
\end{equation}
If we study the creation of $\nu\bar{\nu}$ pairs in a core collapsing supernova (SN) at the bounce stage, one gets that $m_{\mathrm{cr}} \sim 10^{-7}\thinspace\text{eV}$. The obtained upper bound is comparable with the constraint on neutrino masses established earlier in Ref.~\cite{DvoGavGit14}, where we studied the $\nu\bar{\nu}$ pairs creation in SN at the neutronization.

In conclusion we mention that we have studied various phenomena happening with particles electroweakly interacting with accelerated background matter. We have considered two types of the acceleration: due to rotation and a linear acceleration. The exact solutions of the Dirac equation for a test fermion, accounting for both the matter interaction and the noninertial effects, have been found. Then we have discussed the influence of the matter rotation on the resonance in neutrino oscillations, the generation of the electric current flowing along the rotation axis, and the creation of $\nu\bar{\nu}$ pairs in a linearly accelerated matter. Finally we have considered the possibility of the implementation of our results in various astrophysical media such as neutron and quark/hybrid stars as well as SNs.

\section*{Acknowledgments}

I am thankful to the Tomsk State University Competitiveness Improvement Program and to RFBR (research project No.~15-02-00293) for partial support.


\end{document}